\documentclass[aps,prb,twocolumn,superscriptaddress]{revtex4-1}
\usepackage{amsfonts}
\usepackage{graphicx}
\usepackage{epstopdf}
\usepackage{epsfig}
\usepackage{amsmath}
\usepackage[normalem]{ulem}
\usepackage{bm}
\usepackage{color}
\usepackage{makecell}
\usepackage{array}
\usepackage{booktabs}
\usepackage{tabularx}

\newcolumntype{Y}{>{\centering\arraybackslash}X}

\newcommand{\Rmnum}[1]{\expandafter\@slowromancap\romannumeral #1@}
\makeatother

\begin{document}

\title{Decay of semiclassical massless Dirac fermions from integrable and chaotic cavities}

\author{Chen-Di Han}
\affiliation{School of Electrical, Computer and Energy Engineering, Arizona State University, Tempe, Arizona 85287, USA}

\author{Cheng-Zhen Wang}
\affiliation{School of Electrical, Computer and Energy Engineering, Arizona State University, Tempe, Arizona 85287, USA}

\author{Hong-Ya Xu}
\affiliation{School of Electrical, Computer and Energy Engineering, Arizona State University, Tempe, Arizona 85287, USA}

\author{Danhong Huang}
\affiliation{Air Force Research Laboratory, Space Vehicles Directorate, Kirtland Air Force Base, New Mexico 87117, USA}

\author{Ying-Cheng Lai} \email{Ying-Cheng.Lai@asu.edu}
\affiliation{School of Electrical, Computer and Energy Engineering, Arizona State University, Tempe, Arizona 85287, USA}
\affiliation{Department of Physics, Arizona State University, Tempe, Arizona 85287, USA}

\date{\today}

\begin{abstract}

Conventional microlasing of electromagnetic waves requires (1) a high $Q$
cavity and (2) a mechanism for directional emission. Previous theoretical
and experimental work demonstrated that the two requirements can be met
with deformed dielectric cavities that generate chaotic ray dynamics. Is
it possible for a massless Dirac spinor wave in graphene or its photonic
counterpart to exhibit a similar behavior? Intuitively, because of the
absence of backscattering of associated massless spin-1/2 particles and
Klein tunneling, confining the wave in a cavity for a long time seems not
feasible. Deforming the cavity to generate classical chaos would make
confinement even more difficult. Investigating the decay of a spin-1/2 wave
from a scalar potential barrier defined cavity characterized by an effective
refractive index $n$ that depends on the applied potential and the particle
energy, we uncover the striking existence of an interval of the refractive
index in which the average lifetime of the massless spin-1/2 wave in the
cavity can be as high as that of the electromagnetic wave, for both
integrable and chaotic cavities. We also find scaling laws for the ratio
between the mean escape time associated with electromagnetic waves and that
with massless spin-1/2 particles versus the index outside of this interval.
The scaling laws hold regardless of the nature of the classical dynamics.
All the results are verified numerically. The findings provide insights into
the emergent field of Dirac electron optics and have potential applications
in developing unconventional electronics using 2D Dirac materials.


\end{abstract}

\maketitle

\section{Introduction} \label{sec:intro}

Recent years have witnessed vast development of 2D Dirac materials such
as graphene~\cite{Novoselovetal:2004,Novoselovetal:2005,Netoetal:2009},
silicene and germanene~\cite{Wehling:2014,Wang2015}. In these solid state
materials, the energy-momentum relation (dispersion relation) of low-energy
excitations is typically that of a relativistic quantum particle governed by
the Dirac equation. For a massless spin-1/2 Dirac particle, the dispersion
relation is linear which, for the positive energy band, is exactly that of
a photon. It is natural to exploit principles in optics to articulate
strategies to control Dirac electron flows. In this regard, various
optically analogous phenomena such as Fabry-P\'{e}rot
resonances~\cite{Shytov2008,Rickhaus2013}, Talbot effect~\cite{Walls2016}, and
waveguide~\cite{Will2011,Peter2015} in ballistic graphene and similar Dirac
materials have been demonstrated. Due to the negative energy band that
has no counterpart for photons, the nontrivial $\pi$ Berry phase
associated with conical band degeneracy and uniquely relativistic quantum
behaviors such as Klein tunneling~\cite{Klein:1929,Strange:book,KNG:2006}
can arise, leading to unusual physics such as the absence of
backscattering~\cite{ANS:1998,Novikov:2007}, high carrier
mobility~\cite{Geim2007}, and electrically controllable negative refractive
index~\cite{Cheianov2007}. As a result, Dirac particles in ballistic graphene
or other electronic honeycomb lattice crystals can exhibit a number of
unconventional, optical-like behaviors such as negative Goos-H\"{a}nchen
effect~\cite{Beenakker2009}, chirality-assisted electronic
cloaking~\cite{Gu2011}, gate controlled caustics~\cite{Cse2007}, electron
Mie scattering~\cite{Hei2013,Caridad2016,GBKPP:2016,JLee2016}, and whispering
gallery modes~\cite{WF:2014,Zhaoetal:2015,Jiang2017,Ghaharietal:2017}.
Optical-like devices for Dirac particles have also been realized, such as
Klein-tunneling beam splitters and
collimators~\cite{RMRS:2015,Liu2017,VHSWTG:2017} as well as 
microscopes~\cite{Boggild2017}. In addition, the emergent internal degrees
of freedom, i.e., sublattice and valley pseudospins as well as the electron
spin, provide new possibilities for optics based electronic devices such as
valley resolved waveguides~\cite{Wu2011}, beam splitters~\cite{Settnes2016},
electronic birefringent superlenses~\cite{Asm2013}, and spin (current)
lenses~\cite{Moghaddam2010,Zhang2017}. Quite recently, a Dirac quantum chimera
state has been uncovered based on the electronic analog of the chiroptical
effect~\cite{XWHL:2018}. Dirac electron optics~\cite{Cse2007,Cheianov2007,
DSBA:2008,Shytov2008,Beenakker2009,Moghaddam2010,Gu2011,Will2011,Rickhaus2013,
Liao2013,Hei2013,Asm2013,WF:2014,Zhaoetal:2015,Peter2015,Leeetal:2015,
RMRS:2015,Walls2016,Caridad2016,GBKPP:2016,JLee2016,Chenetal:2016,Settnes2016,
Liu2017,VHSWTG:2017,Jiang2017,Ghaharietal:2017,Zhang2017,Boggild2017,
XWHL:2018} have thus become an active field of research.

While optical principles have been exploited in electronics, conventional
optics and photonics have also greatly benefited from the development of
Dirac electronics. For example, the photonic counterparts of Dirac materials
such as graphene, topological insulators and Dirac Semimetals have been
extensively studied, where light is structured in specific ways to mimic
the Dirac particles through the rendering of photonic Dirac cone band
structures. This has led to novel ways to control light with striking
phenomena such as pseudospin-based vortex generation~\cite{Song2015} and
robust light transport~\cite{Lu2014}. Quite recently, inspired by the
emergent topological properties uncovered in gapped Dirac electronic
systems~\cite{Haldane1988}, researchers have made breakthroughs in
topological insulator lasers implemented by topological photonic
cavities~\cite{BWHRSCK:2018,Hararietal:2018}.

Uncovering, understanding, and exploiting the fundamental dynamics of Dirac
particles are thus relevant to both Dirac electronics and photonics. In
this paper, we investigate the trapping of massless Dirac particles in a
scalar potential confinement and the escape from it to address the following
question: is it possible for spin-1/2 Dirac spinor waves in graphene or
photonic graphene systems to exhibit properties similar to those of photons
in a microlasing cavity? To gain insights, we recall the conventional
microlasing systems of electromagnetic waves in a dielectric
cavity~\cite{NSC:1994,MNCSC:1995,NSCGC:1996,NS:1997}, where the geometric
shape of the cavity plays an important role in the wave decay. The dielectric
constant of the cavity is higher than that of the surroundings, so total
internal reflections are responsible for optical ray trapping. For a circular
domain, the classical ray dynamics are integrable, leading to permanent
ray trapping and in principle, to an infinite $Q$ value. However, in
microlasing applications, emission of light is necessary, and one thus
wishes to generate two seemingly contradictory behaviors at the same
time: high $Q$ and good emission directionality. It was theoretically
proposed~\cite{NSC:1994,MNCSC:1995,NSCGC:1996,NS:1997} and experimentally
realized~\cite{GCNNSFSC:1998} that classical chaos can be exploited to
realize both behaviors at the same time, leading to high-$Q$ and
highly efficient microcavity lasing. In nonlinear dynamics, the cavity
problem is closely related to transient chaos~\cite{LL:2002,LT:book} and
leaking~\cite{APT:2013}, with the underlying physics being
non-Hermitian~\cite{CW:2015}. The main advantage that classical chaos can
bring about is that, with simple deformation of the domain boundary, the
phase space is ``mixed'' with coexistence of Kolmogorov-Arnold-Moser (KAM)
tori and chaotic regions, leading to algebraic decay of light rays. The
``long-tail'' nature of the decay gives rise to a high $Q$ value, while the
eventual escape from the chaotic component generates highly directional
emission. In nonlinear dynamics, the exact form of the particle decay law
depends on the relative ``portion'' of the phase space regions whose dynamics
are quasiperiodic (KAM tori) or chaotic. While a mixed phase space gives rise
to algebraic decay with the exponent depending on the amount of domain
deformation, a fully chaotic phase space leads to exponential decay of light
rays. Semiclassically, the cavity problem can be treated by using plane waves
following Fresnel's law, leading to the development of periodic theory of
diffraction~\cite{VWR:1994}, understanding of emission properties in wave
chaotic resonant cavities~\cite{NHJS:1999,Altmann:2009}, uncovering of wave
scars~\cite{RTSCS:2002,LLCMKA:2002,LRRKCK:2004}, analyses of the survival
probability~\cite{RLKP:2006}, directional
emission~\cite{NSCGC:1996,NS:1997,WH:2008}, and Goos-H{\"a}nchen
effect~\cite{SH:2006}. For conventional optics in microlasing
cavities, a general principle is then that the nature of classical dynamics
plays an important role in the decay law.

In order to realize microlasing like behavior, two requirements must be met:
high $Q$ or long lifetime of the wave in the cavity and deformed geometry to
ensure directional emission through classical chaos. Trapping of
massless fermions has recently been experimentally realized in a graphene
confinement~\cite{Zhaoetal:2015,GBKPP:2016,Leeetal:2016,Ghaharietal:2017}.
The geometric shape of the confinement can be chosen to yield classically
integrable, mixed, or chaotic dynamics. We note that, the system is
essentially open with relativistic tunneling defined escape dynamics and thus
generally support trapping modes with a finite lifetime~\cite{HA:2008,
BTB:2009,TOGSM:2010,YHLG:2011a,SB:2011,HSB:2013,SB:2014}. The problem is
also different from that of scattering of Dirac particles from a potential
barrier~\cite{WF:2014,XL:2016}. We focus on the semiclassical regime in which
the plane wave approximation is valid and Fresnel's law is applicable.
Intuitively, due to the total absence of backscattering of massless
spin-1/2 particles~\cite{ANS:1998,Novikov:2007} and the purely relativistic
quantum phenomenon of Klein tunneling~\cite{Klein:1929,Strange:book,KNG:2006},
the decay of the spinor wave would be enhanced when comparing with that of
classical electromagnetic waves from the same cavity, so trapping of the
former would seem impossible. Indeed, a detailed scaling analysis of the
ratio of the mean escape time of an electromagnetic wave to that of a spin-1/2
wave reveals that, for both integrable and chaotic cavities, in the regime
of large effective refractive index values ($n \gg 1$), the ratio is
proportional to $n$ but in the regime of $n \ll 1$, the ratio is inversely
proportional to $n$. This means that, in these two asymptotic regimes, the
averaging lifetime of the spin-1/2 wave is indeed much smaller than that
of the electromagnetic wave. The surprising phenomenon is that, in between
the two asymptotic regimes, an interval in $n$ emerges, in which the ratio
is about one, indicating that the spin-1/2 wave can live as long as the
electromagnetic counterpart. This means that, high $Q$ can be achieved for
spin-1/2 particles. Since the constant ratio also holds for classical
chaotic cavities, nonisotropic coherent emission can be
expected. The finding suggests strongly that the two microlasing conditions
for photons can be fulfilled for spin-1/2 particles. Our analysis provides
insights into Dirac electronics and photonics, and has potential applications
in developing unconventional cavity laser designs based on Dirac photonic
crystals, and optical-like electronics with 2D Dirac materials.

We remark that, for photonic graphene 
systems~\cite{ZL:2008,ZdeD:2010,BDMORS:2010},
the concept of lasing can be defined since the underlying particles are
actually photons. For Dirac fermions, an analog is atom laser~\cite{BHE:1999},
that emits a beam of atoms (not light). Rigorously, the concept of 
``lasing'' does not hold for an atom laser. In fact, traditional atom 
lasers require Bose Einstein condensate (BEC), although there was an  
attempt to generate an atom laser without BEC~\cite{RLCWG:2006}.
In this paper, the term ``lasing'' is loosely used for photonic graphene 
systems. For Dirac fermions, we use the term ``coherent emission.''

\section{Ray dynamics for spin-1/2 fermion} \label{sec:ray_tracing}

We focus on the semiclassical regime where the wavelength of the particle
is relatively small in comparison with the size of the system but is
non-negligible. In the semiclassical regime, both quantum and classical
behaviors are relevant, and it is the ideal regime to study the quantum
manifestations of distinct types of classical dynamics including chaos.
In fact, most previous work in the traditional field of (nonrelativistic)
quantum chaos~\cite{Stockmann:book,Haake:book} emphasizes the importance of
the semiclassical regime.

\begin{figure*}
\centering
\includegraphics[width=\linewidth]{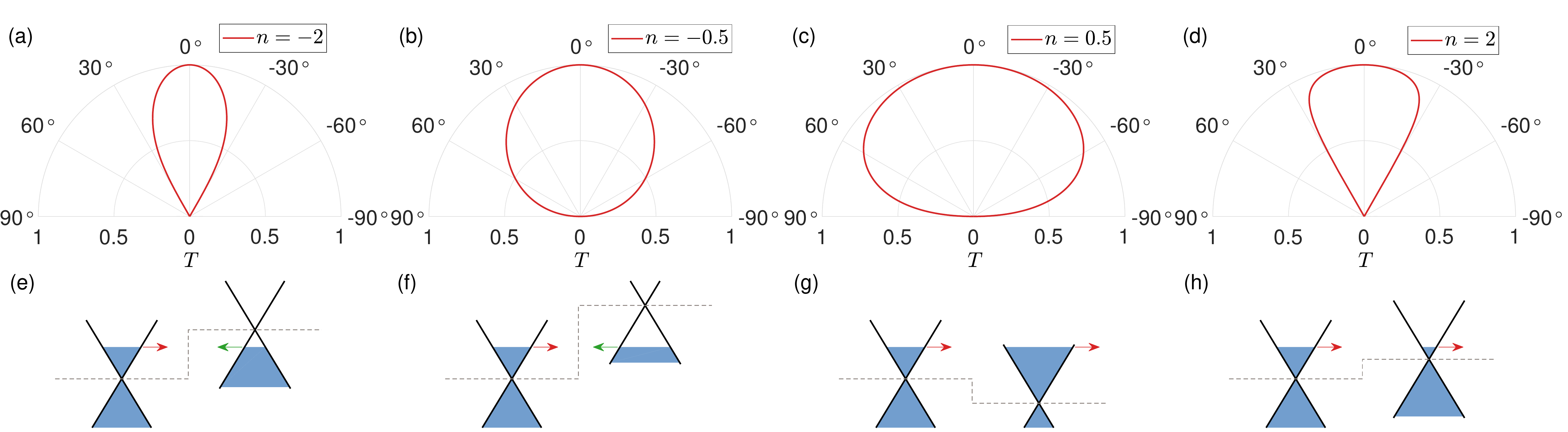}
\caption{ {\bf Approach of Dirac electron optics to solving the fermion
decay problem: four distinct intervals of the effective refractive index in
the domain of electrical potential confinement}. (a-d) The transmission
coefficient versus the incident angle in the polar representation for
$n \in (-\infty, -1]$, $n \in [-1,0]$, $n\in [0,1]$, and $n \in [1,\infty)$,
respectively. (e-h) The corresponding Dirac cone structures outside and
inside of the confinement region. Note that, for (a,b) [or (e,f)], inside
the potential region, the directions of the wavevector and the velocity
are opposite to each other because of the negative refractive index in these
two cases.}
\label{fig:T_four_regions}
\end{figure*}

Trapping of a spin-1/2 fermion can be realized through a confinement (cavity)
of electrical potential given by~\cite{GBKPP:2016}
\begin{equation}
V(\mathbf{r})=\begin{cases}
0, &\mathbf{r} \in  \mathcal{D}, \\
V_0, &\mathbf{r}  \notin  \mathcal{D}, \\
\end{cases}
\end{equation}
where $\mathcal{D}$ represents the geometrical domain of the cavity and
$V_0$ is the uniform potential applied to the domain.
Classically, the cavity is equivalent to a billiard, where the ray behavior
is identical to that of a point particle bouncing back and forth in the
billiard, with the difference that an optical ray is subject to reflection
and refraction. The geometry shape of $\mathcal{D}$ can then be chosen
to generate characteristically distinct types of classical dynamics: from
integrable to fully chaotic. 
To be concrete, in this paper, we focus on two types of geometrical
shapes for the cavity: a circle or a square that generates classically 
integrable dynamics and a stadium in which the classical dynamics are 
chaotic. Experimentally, for a 2D solid state material
(e.g., graphene), the domain $\mathcal{D}$ can be realized through the
technique of scanning tunneling microscope
(STM)~\cite{Zhaoetal:2015,GBKPP:2016,Leeetal:2016,Ghaharietal:2017}.

The traditional theoretical approach consists of writing down the
non-Hermitian Hamiltonian and solving the Dirac equation subject to proper
boundary conditions~\cite{HA:2008}. If the domain shape is simple and
highly symmetric, e.g., a circle, which yields classically integrable
dynamics, then the solutions of the Dirac equation can be readily obtained.
When the boundaries of the domain are deformed from the circular
shape to generate chaotic dynamics, if the boundary conditions are of the
infinite mass confinement type, numerical solutions of the Dirac equation can
be obtained using the boundary integral method~\cite{BM:1987} or the standard
finite element algorithm~\cite{NHLG:2012}. In our problem of particle trapping
and decay, the boundary condition is not of the infinite mass confinement type.
In this case, for a domain of an arbitrary shape, even numerical solutions
of the Dirac equation are extremely difficult. Since our focus is on the
semiclassical regime, we take advantage of the field of Dirac electron optics
to solve the Dirac equation by using the approach of ray tracing associated
with conventional wave optics.

When an electromagnetic wave encounters a boundary, reflection and refraction
occur as governed by Fresnel's law. In the underlying ray picture, there
will be energy loss associated with each encounter with the domain boundary.
For a spin-1/2 fermion, Klein tunneling must be taken into account to derive
the corresponding Fresnel's law. Depending on the particle energy $E$
relative to the potential height $V_0$, there are two distinct
cases~\cite{AF:2011,OABC:2017}: (i) $0<E<V_0$ and (ii) $V_0<E$. In the first
case ($0<E<V_0$), the transmission and reflection coefficients, $T$ and $R$,
respectively, at each encounter with the boundary are given
by~\cite{AF:2011,OABC:2017}
\begin{eqnarray} \label{eq:case_1_minus}
T & = & -\frac{2\cos{\theta}\cos{\theta_t}}{1-\cos{(\theta+\theta_t)}},
\\ \nonumber
R & = & \frac{1+\cos{(\theta-\theta_t)}}{1-\cos{(\theta+\theta_t)}},
\end{eqnarray}
where $\theta$ is the incident angle with respect to the normal and the
transmitted angle is given by
\begin{equation} \label{eq:case_1_theta_t}
\theta_t = \sin^{-1}{\left(\frac{E}{E-V_0}\sin{\theta}\right)}+\pi
=\sin^{-1}{(n\sin{\theta})} + \pi.
\end{equation}
with the effective refractive index $n$ defined as $n \equiv E/(E-V_0)$, which
is negative in this case: $n\in (-\infty, 0]$.

For the second case $V_0<E$, the transmission and reflection coefficients
are given by
\begin{eqnarray} \label{eq:case_1_plus}
T & = & \frac{2\cos{\theta}\cos{\theta_t}}{1+\cos{(\theta+\theta_t)}},
\\ \nonumber
R & = & \frac{1-\cos{(\theta-\theta_t)}}{1+\cos{(\theta+\theta_t)}},
\end{eqnarray}
where the refracted angle is
\begin{equation}
\theta_t = \sin^{-1}{(n\sin{\theta})},
\end{equation}
and the effective refractive index is positive: $n\in [0,\infty)$.

The energy band structure associated with a spin-1/2 fermion is that of
a pair of Dirac cones. Depending on the relative positions of the Dirac cone
structures outside and inside of the potential domain, there are four
distinct intervals of the refractive index: $(-\infty, -1]$, $[-1,0]$,
$[0,1]$ and $[1,\infty)$. Figures~\ref{fig:T_four_regions}(a-d) show the
transmission coefficient versus the incident angle in the polar
representation for the four parameter intervals, respectively. The
corresponding energy band structures are shown in
Figs.~\ref{fig:T_four_regions}(e-h), respectively.

The survival probability of a spin-1/2 fermion inside the potential region
can be calculated using the formulas for the transmission and reflection
coefficients. In general, the coefficients in the Klein tunneling regime 
depend on the incident energy $E$ and the potential height $V_0$ (which
together define the effective refractive index $n$~\cite{AF:2011}) as well as 
the angle of incidence $\theta$. There is a symmetry in the coefficients in 
that they do not depend on the sign of $n$, which can be seen by substituting 
the expression of $\theta_t$ into Eq.~\eqref{eq:case_1_minus} or 
Eq.~\eqref{eq:case_1_plus}. It thus suffices to focus on the two distinct 
intervals of the values of the refractive index: $|n|<1$ and $|n|>1$.
For $|n|>1$, total internal reflection can occur with the critical incident 
angle of $\theta_c$, where the transmission coefficient is zero for 
$\theta > \theta_c$.

\section{Results} \label{sec:result}

Say we distribute an ensemble of rays of spin-1/2 waves with different
initial conditions in the cavity. As a ray evolves following Fresnel's law,
its intensity will decrease due to refraction. Let $I_0$ be the initial
intensity (or energy) of any ray in the ensemble. After $n$ encounters with
the boundary, the intensity becomes
\begin{equation} \label{eq:2_remaining_strength}
I_n=I_0 \prod_i R_i=I_0 \exp \left(\sum_i \ln R_i \right).
\end{equation}
The survival probability $P(t)$ is the fraction of the remaining intensity
at time $t$. Depending on the nature of the classical ray dynamics (integrable
or chaotic) and on the value of the effective refractive index $n$, with
time $P(t)$ decays either exponentially:
\begin{equation}
P(t) = \exp(-\gamma t),
\end{equation}
where $\gamma$ is the exponential decay rate, or algebraically:
\begin{equation}
P(t) \sim t^{-z},
\end{equation}
with $z$ being the algebraic decay exponent.

In numerical simulations, we initialize a large number of rays (between
$10^5$ and $10^7$) randomly distributed on the boundary. For each ray, the
initial angle $\theta_0$ is chosen according to $p=\sin{\theta_0}$, where
$p$ is a uniform random variable in the unit interval and the velocity is
chosen to be one. The final distribution of the rays is independent of the
initial random conditions~\cite{LT:book}. In the following, we treat
integrable and chaotic cavities separately.

\subsection{Integrable cavity} \label{subsec:integrable_cavity}

We consider a circular cavity with integrable ray dynamics, in which
the incident angle $\theta$ is constant and the time interval between two
successive encounters with the boundary is $\Delta t = 2\cos{(\theta)}$.
For $|n|<1$, the ray intensity decays exponentially. As indicated in
Fig.~\ref{fig:T_four_regions}, the transmission coefficient $T$ takes on
the minimum value at $\theta=\pi/2$. A ray with $\theta=\pi/2$ can thus
survive in the cavity for a long time.

\begin{figure}
\centering
\includegraphics[width=\linewidth]{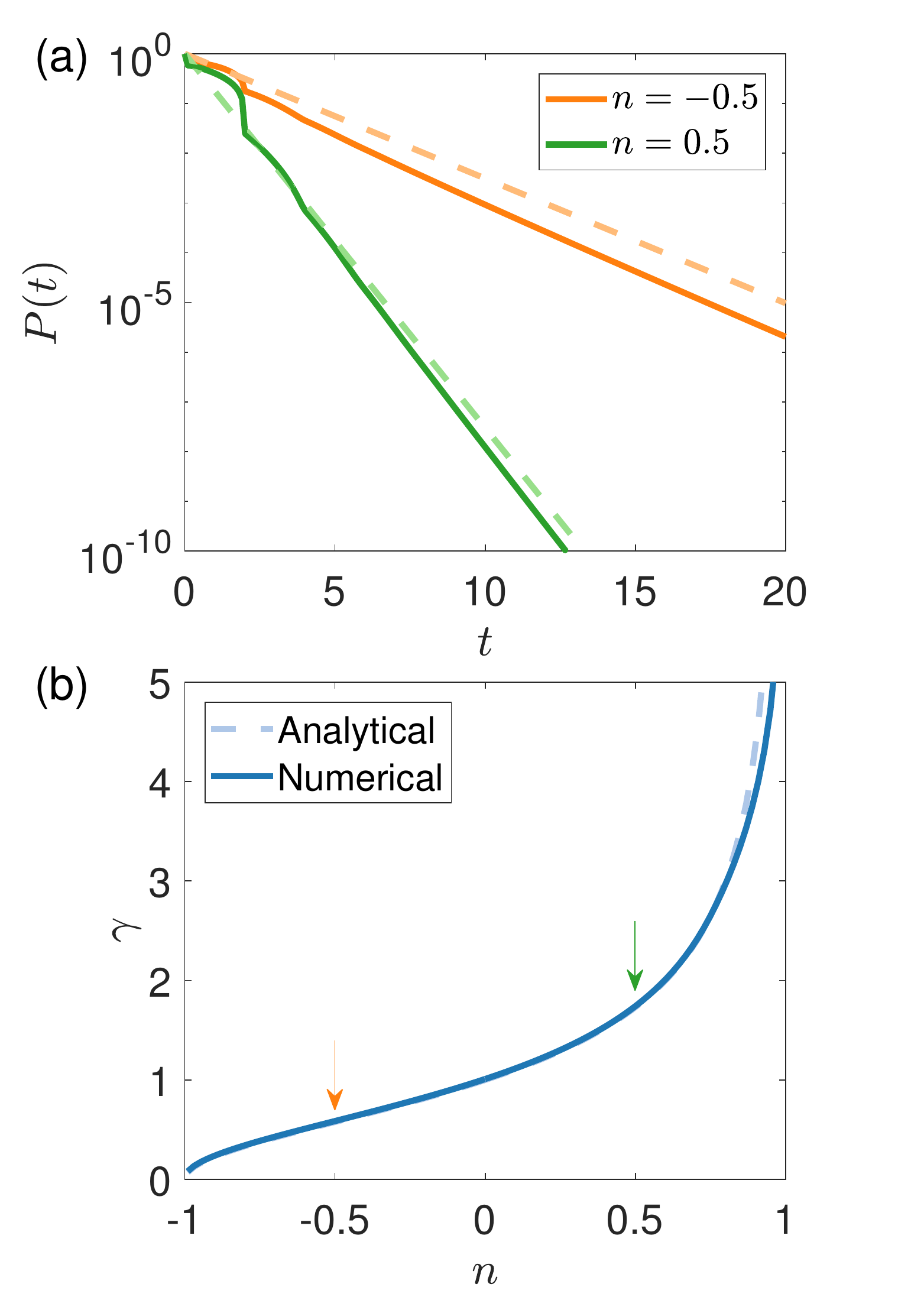}
\caption{ {\bf Exponential decay of the survival probability for the integrable
(circular) cavity for $-1 < n < 1$}. (a) Two examples of exponential decay
of $P(t)$ ($n = -0.5$ and $n = 0.5$). (b) The exponential decay exponent
$\gamma$ versus $n$. Insofar as the value of $n$ is not close to one, the
whispering gallery mode decays most slowly. For $n \rightarrow 1$, the cavity
becomes transparent so $\gamma \rightarrow \infty$.}
\label{fig:circle_gamma}
\end{figure}

For $0<n<1$, the survival probability is given by
\begin{equation} \label{eq:2_circle_1_1_calculation}
P(t) = \frac{\sum_i R_i^{t/\Delta t}}{N} =
\lim_{\theta\rightarrow \pi/2} \left(\frac{1-\cos(\theta-\theta_t)}
{1+\cos(\theta+\theta_t)} \right)^{t/(2\cos\theta)}.
\end{equation}
Letting $x=\frac{\pi}{2}-\theta$, we expand $P(t)$ at $x=0$ to obtain
\begin{equation} \label{eq:2_circle_1_1}
P(t) \approx \exp\left(-\frac{\sqrt{1-n^2}}{1-n}t \right).
\end{equation}
The survival probability for $-1<n<0$ has the same exponential form but
with different values of the decay exponent.
Figure~\ref{fig:circle_gamma}(a) shows the decay of the survival probability
$P(t)$ with time for $n = -0.5$ and $n = 0.5$ on a semi-logarithmic scale.
Each set of data can be well fitted by a straight line, indicating that the
decay is exponential. In both cases, the mode that survives the longest
possible time in the cavity is the whispering gallery mode with
$\theta \alt \pi/2$, and this holds insofar as the value of $n$ is not too
close to one. Figure~\ref{fig:circle_gamma}(b) shows, for $-1 < n < 1$, the
exponential decay exponent $\gamma$ versus $n$. For $n\rightarrow 1$, the
electrical potential vanishes, so the cavity becomes transparent, resulting
in infinitely fast decay, i.e., $\gamma \rightarrow \infty$.

\begin{figure}
\centering
\includegraphics[width=\linewidth]{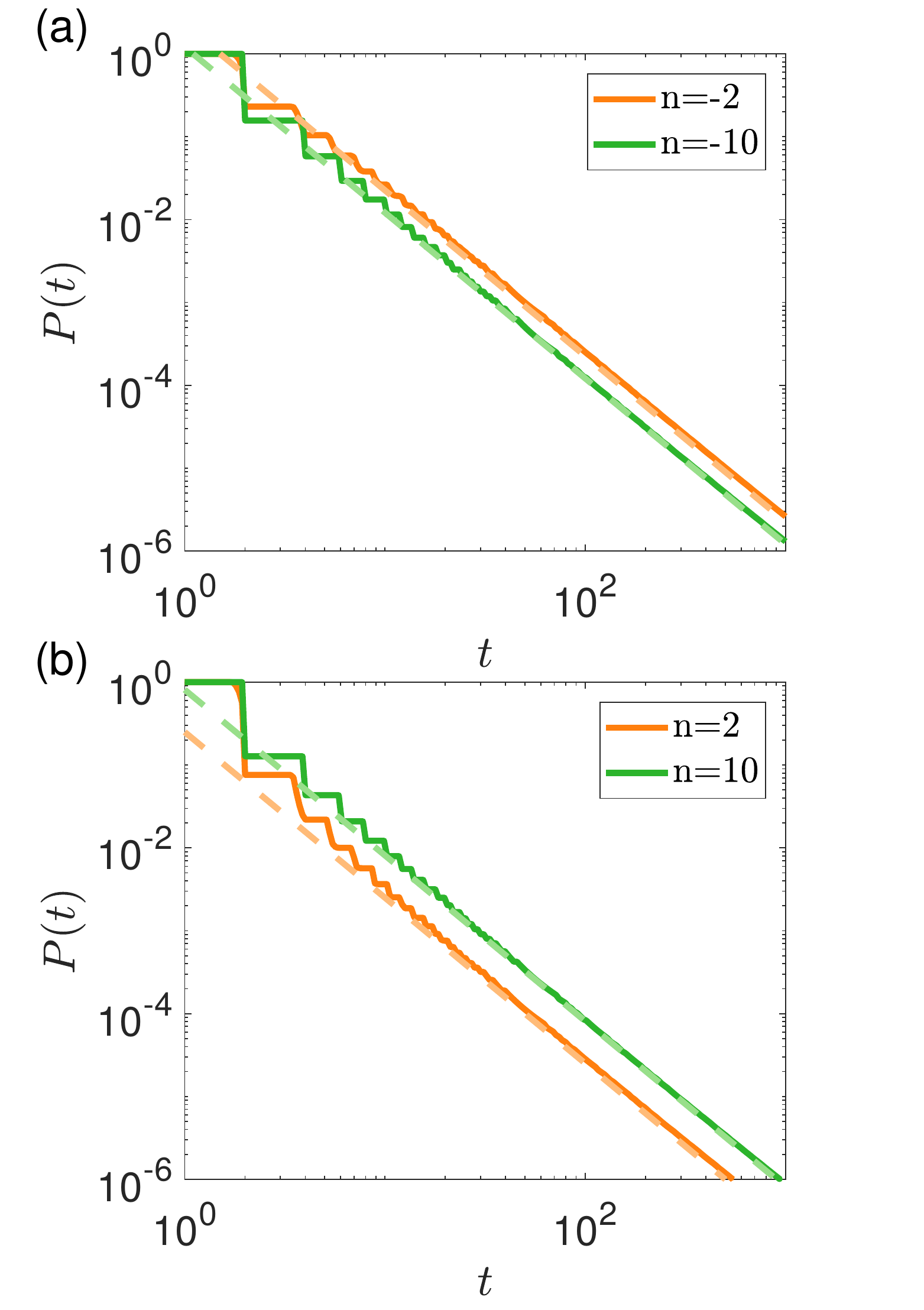}
\caption{ {\bf Algebraic decay of the survival probability for the integrable
(circular) cavity for $|n| > 1$}. Shown are decay behaviors of $P(t)$ for
(a) $n=-2$ and $n=-10$ and (b) $n=2$ and $n=10$. The solid and dashed lines
represent numerical and theoretical results, respectively.}
\label{fig:circle_alge}
\end{figure}

For $n > 1$, total internal reflections occur for
$\theta > \theta_c = \sin^{-1}{(1/n)}$. In this case, the modes that can survive
for a long time are those with incident angle near $\theta_c$, and $P(t)$ is
given by
\begin{equation} \label{eq:P_integral}
P(t) = \frac{\int_0^{\theta_c} \cos{\theta}
\exp{\left[ -G(\theta)t \right]} d\theta }{\int_0^{1/n} dp},
\end{equation}
where
\begin{equation}
G(\theta)=\frac{1}{2\cos{\theta}}
\ln{\left(1+\frac{2\cos{\theta}\cos{\theta_t}}
{1-\cos{(\theta-\theta_t)}} \right)}.
\end{equation}
Letting $x = \theta_c - \theta$, we have
\begin{equation} \label{eq:2_circle_1_inf_expansion}
G(\theta)\approx \alpha \sqrt{x}+O(x).
\end{equation}
where
\begin{equation}
\alpha=\frac{n\sqrt{2\sqrt{n^2-1}}}{n-1}.
\end{equation}
Substituting Eq.~(\ref{eq:2_circle_1_inf_expansion}) into
Eq.~(\ref{eq:P_integral}), we get
\begin{eqnarray}
\nonumber
P(t) & \approx & \frac{2\sqrt{n^2-1} }{\alpha^2 t^2}
[1-(1+\alpha\sqrt{\theta_c}t)e^{-\alpha \sqrt{\theta_c} t}] \\
& \approx & \frac{2\sqrt{n^2-1}}{\alpha^2} t^{-2}.
\label{eq:2_circle_1_inf}
\end{eqnarray}
We thus see that the decay of $P(t)$ for $n>1$ is algebraic. A similar
analysis gives that Eq.~(\ref{eq:2_circle_1_inf}) holds for the $n<-1$
region. Numerical validation of Eq.~(\ref{eq:2_circle_1_inf}) is given
in Fig.~\ref{fig:circle_alge}.

The special parameter point $n=1$ is one at which the decay law changes
characteristically from being exponential to being algebraic. This is not
surprising because of the emergence of total internal reflections for $n > 1$.
Another special parameter value is $n = -1$, which occurs when the particle
energy is one half of the potential height: $E=V/2$. In this case, the
transmission coefficient at a single encounter with the cavity boundary is
$T=\cos^2{\theta}$, so the critical incident angle is $\theta_c=\pi/2$.
Near $\theta_c$, the quantity $n\sin\theta_c$ can no longer be treated as
a small number, so the expansion used in deriving Eq.~(\ref{eq:2_circle_1_inf}) is not valid. However, we can rewrite $G(\theta)$ as
\begin{equation}
G(\theta)=-\frac{1}{2\cos\theta}\ln (1-\cos^2\theta).
\end{equation}
For $\theta\rightarrow \pi/2$, we have $\cos{\theta} \sim x$ and obtain
\begin{equation}
P(t)\approx \int_0^{\pi/2} x\exp\left(-\frac{tx}{2}\right) d\theta
=\frac{4}{t^2},
\end{equation}
which is valid in the large $t$ regime.

\subsection{Effect of Klein tunneling on decay of spin-1/2 wave}
\label{subsec:KT}

In terms of the cavity decay dynamics, what is the key difference between
an electromagnetic wave and a Dirac spinor wave? 
For a Dirac particle, there
is a fundamental phenomenon that has no counterpart for a photon: Klein
tunneling~\cite{Klein:1929,Strange:book}, a uniquely relativistic quantum
phenomenon by which a particle of energy less than the height of a potential
barrier can tunnel through it with absolute certainty. 
For a Dirac electron optical system, Klein tunneling occurs in the $|n|< 1$
regime for $\theta = 0$ because, from Eq.~(\ref{eq:case_1_minus}), we have
$T(\theta=0)=1$. In this regime, the decay of both spin-1/2 and electromagnetic
waves is exponential (see Table~\ref{tab:time_scale} below). However, the
slowest decaying modes are quite different for the two types of waves. In
particular, for the spin-1/2 wave they are the whispering gallery  modes
(corresponding to $\theta \approx \pi/2$) as it is difficult for the modes
with $\theta \approx 0$ to stay in the cavity for a long time because of Klein
tunneling. For the electromagnetic wave, the situation is nearly opposite:
the longest survival modes are those with $\theta$ near zero, i.e., modes
with propagation along the diameter of the cavity, as the transmission
coefficients are minimum for them by Fresnel's law.

\begin{figure}
\centering
\includegraphics[width=\linewidth]{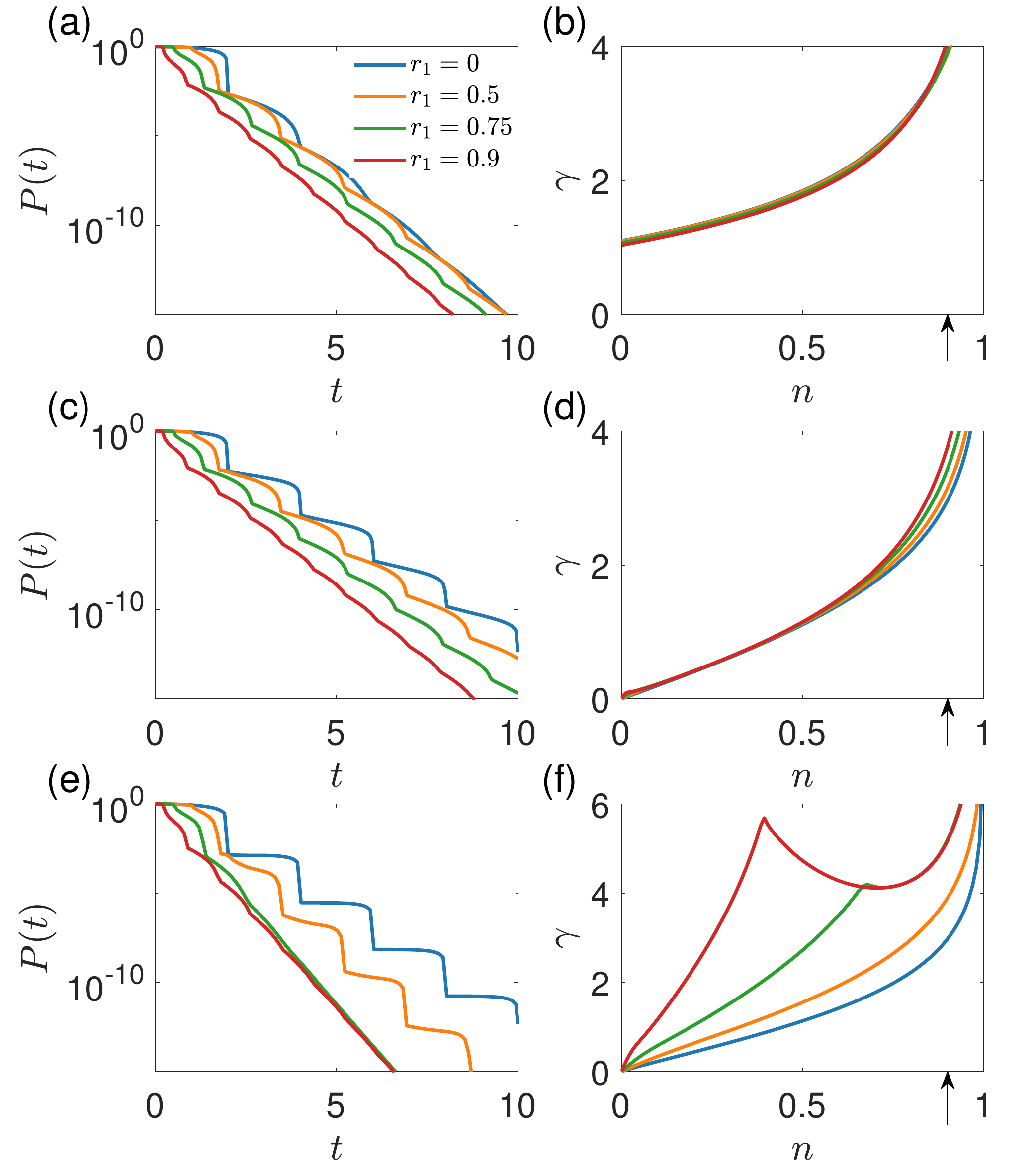}
\caption{ {\bf Decay of spin-1/2 and electromagnetic waves from a ring
cavity}. (a) For four different ring configurations ($r_1 = 0$ - the
original circular cavity, $r_1 = 0.5$, 0.75, and 0.9), exponential decay
behavior of the survival probability $P(t)$, and (b) the corresponding
decay exponent $\gamma$ versus $n$ for $0 < n < 1$. As the long time decay
behavior is dominated by the whispering gallery modes that concentrate on
the larger circumstance, the central forbidden region has little effect on
the overall exponential decaying behavior. (c,d) The corresponding results
for a TM electromagnetic wave, where the decay rates for the three actual
ring configurations are larger than that of the original circular cavity,
due to the blockage of the slowest decaying modes along the diameter.
(e,f) Results for a TE electromagnetic wave, where the inaccessibility of
the central region has an even more significant impact on the decay.}
\label{fig:ring}
\end{figure}

To further appreciate the difference between the decay dynamics of spin-1/2
and electromagnetic waves, we study the ring cavity with a kind
of a small ``forbidden'' region at the center of the circular cavity
($r < r_1 < 1$). The basic idea is that, the presence of the forbidden region
should not have a significant impact on the decay of a spin-1/2 wave as the
dominant surviving modes are of the whispering gallery type, which do not
pass through the central region of the cavity. However, the forbidden region
would affect the decay of the electromagnetic wave as the modes that
shape the decay behavior are diametrical, which have a significant presence
in the central region. The ring cavity is defined as
\begin{equation} \label{eq:3_ring_n}
n(r)=\begin{cases}
\infty, & r \leq r_1, \\
(0,1), & r_1 < r \leq 1, \\
1, & r > 1,
\end{cases}
\end{equation}
which is integrable. In terms of the ray dynamics, the central circular
region blocks most orbits along the diameter. Figure~\ref{fig:ring}(a) shows,
for spin-1/2 wave and $n = 0.9$, the exponential decay of the survival
probability $P(t)$ for four different ring configurations (corresponding
to different values of $r_1$) on a semi-logarithmic
scale. The decay curves can be fitted approximately by lines with nearly
identical slopes, indicating that introducing a central forbidden region has
little effect on the decay. Figure~\ref{fig:ring}(b) displays the curves of
the exponential decay exponent $\gamma$ versus $n$ (for $0<n<1$) for the four
different ring configurations, which are nearly identical.
Figures~\ref{fig:ring}(c) and \ref{fig:ring}(d)
present the corresponding results for a TM electromagnetic
wave (Appendix~\ref{Appendix_A}), revealing a significant effect of
the central forbidden region on the wave decay behavior. In particular,
the decay rates for the three cases of $r_1 \ne 0$ (orange, green and
red curves) are greater than that of the circular cavity ($r_1 = 0$, the blue
curve) because the central forbidden region blocks the slowest decaying
modes so as to expedite the overall decay. The results for a TE electromagnetic
wave are shown in Figs.~\ref{fig:ring}(c) and \ref{fig:ring}(d), revealing
an even more significant effect of the blockage of the central region on
wave decay.

Experimentally, for a spin 1/2 wave system, e.g., graphene, the central
circular region can be created by applying an electrical potential
corresponding exactly to the Fermi energy $E$. For the electromagnetic
wave, the ring configuration can be realized by depositing metal in the
central region of a circular dielectric cavity, which induces total
internal reflections.

\subsection{Chaotic cavity} \label{subsec:chaotic_cavity}

\begin{figure}
\centering
\includegraphics[width=\linewidth]{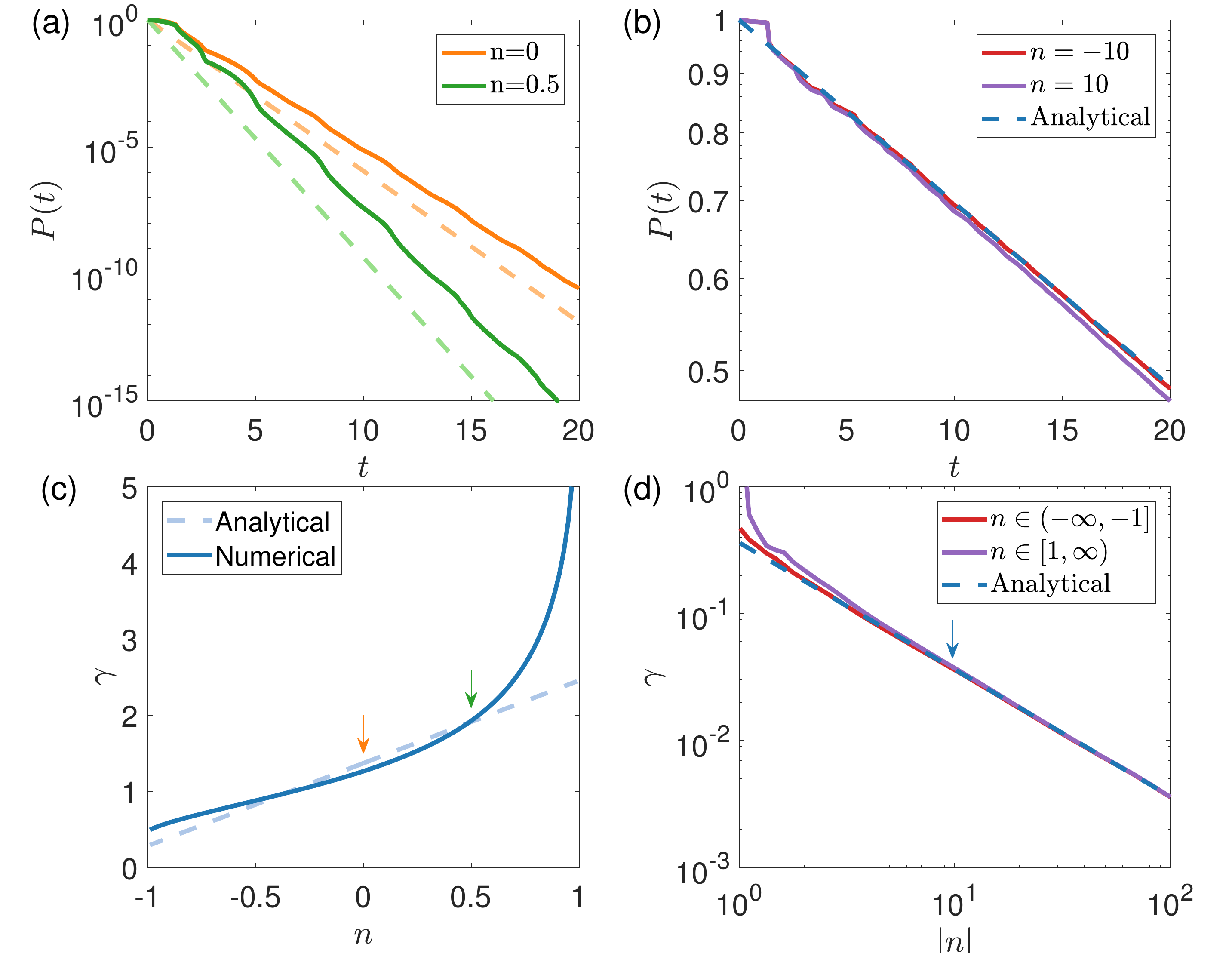}
\caption{ {\bf Exponential decay of a spin-1/2 fermion from a chaotic
stadium cavity}. (a) The survival probability $P(t)$ for $n=0$ and $n=0.5$
(on a semi-logarithmic scale), where the dashed lines are theoretical
predictions. (b) Exponential decay of $P(t)$ for $n=-10$ and $n=10$.
(c) The exponential decay exponent $\gamma$ versus $n$ for $-1 < n < 1$.
(d) The exponent $\gamma$v versus $|n|$ for $1 < |n| < 100$. In (c) and
(d), the dashed line is the analytic prediction. The parameters of the
chaotic stadium are $A = \pi$, $r_0/l = 1/2$, and $L = 2\pi r_0 + 2l$,
where $l$ is the length of the straight segment.}
\label{fig:decay_chaotic_cavity}
\end{figure}

To be concrete, we consider the chaotic stadium cavity characterized
by parameters $r_0$ (the radius of each semicircle) and $L$ (the perimeter of
the whole domain). It is useful to define~\cite{MLS:1993} the ``average path
length'' $\langle d\rangle=\pi A/L$, where $A$ is the area of the stadium.
The survival probability is given by
\begin{equation} \label{eq:2_stadium_summation}
P(t)=\prod_i R_i=\exp{\left[\frac{t}{\langle d\rangle}\sum_i\ln{R_i}\right]},
\end{equation}
where $R_i$ is the reflection coefficient at each encounter with the boundary.
The summation can be approximated by a double integral in both distance and
angle. Due to chaos, the distance between two successive encounters with
the boundary is roughly constant ($\langle d\rangle$), so the double
integral can be reduced to a single integral with respect to the angle. The
summation in Eq.\eqref{eq:2_stadium_summation} can then be evaluated as
\begin{equation} \label{eq:2_stadium_integration}
\frac{\int_0^{\pi/2} \cos{\theta} \ln{R} d\theta}
{\int_0^{\pi/2} \cos{\theta} d\theta}.
\end{equation}
In general, the integral cannot be evaluated analytically. However, in the
limiting cases ($|n|\rightarrow 0$ and $|n|\rightarrow \infty$), we can use
Taylor expansion to evaluate the integral. The end result is an exponential
decay of $P(t)$ (Appendix~\ref{Appendix_B}) with explicit formulas for the
decay exponent $\gamma$. In particular, for $n\rightarrow 0$, $\gamma$
is given by
\begin{equation}
\gamma=\frac{1}{\langle d\rangle}\left(\sum_{i=1}^\infty
\frac{W_{2i+1}}{i} + \frac{\pi}{2} n\right),
\end{equation}
where $W$ is the Walli's integral (Appendix~\ref{Appendix_B}). For
$n\rightarrow\infty$, we have
\begin{equation}
\gamma=\frac{(8-2\pi)\pi r_0}{\langle d\rangle |n|L}.
\end{equation}
Numerical verification is presented in Figs.~\ref{fig:decay_chaotic_cavity}.
These results suggest that the decay of a spin-1/2 fermion from a chaotic
cavity is generally exponential, implying the difficulty in confining the
relativistic quantum particle.

\subsection{Are high $Q$ and nonisotropic coherent emission 
achievable for a spin-1/2 wave?}

Two conditions must be met in order for a microcavity to generate effective
lasing: (1) high $Q$ value, and (2) directional emission.
Previous work on conventional electromagnetic
microlasing~\cite{NSC:1994,MNCSC:1995,NSCGC:1996,NS:1997,GCNNSFSC:1998}
established that deformed chaotic cavities are suited for microlasing
applications. To realize microlasing of a spin-1/2 wave, it is necessary
that (1) the (graphene) cavity has average lifetime comparable to that of
the dielectric electromagnetic cavity, and (2) the lifetime can be
maintained in a deformed chaotic cavity. In the following, we establish
the existence of a range of the effective refractive index value for
a spin-1/2 particle cavity in which the two requirements can be met.

To compare the decay of the spin-1/2 wave with that of the electromagnetic wave
in the same cavity, it is necessary to have a complete picture of the decay
of the electromagnetic wave from a cavity for comparison. Especially, for
a spin-1/2 particles, in principle the relative refractive index $n$ can
take on values ranging from $-\infty$ to $+\infty$.
For electromagnetic waves, previous work~\cite{RLKP:2006} treated this
problem but for the case where the absolute value $|n|$ of the relative
refractive index of the dielectric cavity is greater than one, with the
result that the decay law is algebraic (exponential) for integrable (chaotic)
cavities. As a necessary step, we extend the result to the $|n|<1$ regime
(Appendix~\ref{Appendix_A}). Table~\ref{tab:time_scale} lists the formulas
of $P(t)$ for both spin-1/2 and electromagnetic waves (TE and TM) for
both integrable and chaotic cavities.

\begin{table*}
\caption{ {\bf Complete results of the survival probability for spin-1/2,
TE and TM electromagnetic waves for integrable and chaotic cavities}.
For algebraic decay, the exact form of $P(t)$ is listed. For exponential
decay, only the decay exponent $\gamma$ is given.}
\label{tab:time_scale}
\begin{tabularx}{\textwidth}{YYYYY}
\hline\hline
\specialrule{0em}{1pt}{1pt}
& \multicolumn{2}{c}{Circular cavity} & \multicolumn{2}{c}{Stadium cavity} \\
& $ |n|<1$ & $|n|>1$ & $ |n|<1$ & $|n|>1$ \\
\specialrule{0em}{1pt}{1pt}
\hline
\specialrule{0em}{2pt}{2pt}
Spin 1/2  & $\gamma= \frac{\sqrt{1-n^2}}{1-n} $ & $P(t)=\frac{(n-1)^2}{n^2} t^{-2}$& $\gamma=\sum_i \frac{W_{2i+1}}{\langle d \rangle i}+ \frac{\pi n}{2\langle d \rangle} $ & $\gamma=\frac{(8-2\pi)\pi r_0}{\langle d\rangle |n| L}$  \\
\specialrule{0em}{2pt}{2pt}
TM &$\gamma=\ln\left(\frac{1+n}{1-n} \right)$ & $P(t)=\frac{(n^2-1)^2}{4n^2}t^{-2}$ & $\gamma=\frac{n\pi}{\langle d \rangle}$  &  $\gamma=\frac{2 \pi^2 r_0}{\langle d\rangle L n^2}$ \\
\specialrule{0em}{2pt}{2pt}
TE  & $\gamma=\ln\left(\frac{1+n}{1-n} \right) $ & $\gamma=\ln\left(\frac{n+1}{n-1} \right)$ & $\gamma=\frac{2n\pi}{\langle d \rangle}$ & $\gamma=\frac{4 \pi^2 r_0}{\langle d\rangle L n^2}$ \\
\specialrule{0em}{2pt}{2pt}
\hline\hline
\end{tabularx}
\end{table*}

For the cavity decay problem, a basic characterizing quantity is the quality
factor $Q$, which qualitatively measures the stability of the wave
(temporarily) ``trapped'' in the cavity. To calculate the $Q$ value, we
resort to the fact that, because the system is fundamentally open, the
underlying Hamiltonian is non-Hermitian with complex eigenvalues, and $Q$
is nothing but the ratio between the real and imaginary parts of the
complex eigen wave vector. Alternatively, $Q$ can be defined as
$Q \equiv \omega_n \tau$, where $\omega$ is the frequency of the dominantly
surviving mode, $\tau$ is the associated (finite) lifetime, and its inverse
is the spectral width~\cite{CW:2015}. In the ray picture, it is convenient
to calculate the mean escape time (or lifetime) $T$, which is the time
for $P(t)$ to reduce to the value of, e.g., $e^{-1}$.

Intuitively, because of Klein tunneling, it would be ``easier'' for a spin-1/2
wave to leak out of the cavity than an electromagnetic wave. Let
$\tau_{\mbox{\tiny S}}$, $\tau_{\mbox{\tiny TE}}$, and $\tau_{\mbox{\tiny TM}}$
denote the mean escape time for spin-1/2, TE and TM electromagnetic waves,
respectively. We analyze the ratios  $\tau_{\mbox{\tiny TE}}/\tau_{\mbox{\tiny S}}$ and
$\tau_{\mbox{\tiny TM}}/\tau_{\mbox{\tiny S}}$, which can be calculated based
on the results in Table~\ref{tab:time_scale}.

For the $|n| \ll 1$ regime, all three systems exhibit exponential decay, so we
have $\tau = 1/\gamma$. For electromagnetic waves, the average lifetime is
proportional to $1/n$ but for the spin-1/2 system the time tends to a
constant. We thus have
\begin{equation}
\tau_{\mbox{\tiny EM}}/\tau_{\mbox{\tiny S}} \sim \frac{1}{n} \gg 1 \ \text{for} \ n\rightarrow 0,
\end{equation}
for the integrable cavity, where $\tau_{\mbox{\tiny EM}}$ stands for either $\tau_{\mbox{\tiny TE}}$ or
$\tau_{\mbox{\tiny TM}}$. The same result holds for the chaotic cavity. In this case,
comparing with the electromagnetic wave, a spin-1/2 wave will leak out of
the cavity more quickly, i.e., it is less ``stable'' when being compared
with the electromagnetic wave.

In the $|n| \gg 1$ regime, for the integrable cavity, we have
$P(t)\sim n^2 t^{-2}$ for the TM electromagnetic wave, so the mean escape
time $T_{\mbox{\tiny TM}}$ is proportional to $n$. For the TE wave, the decay
is exponential with the exponent given by $\gamma=\ln{(n+1)/(n-1)}$. We thus
have $\gamma \approx 2/n$ and, hence, $\tau_{\mbox{\tiny TE}} \sim n$, as for
the TM wave. For the chaotic cavity, for both TM and TE waves, we have
$\gamma\sim n^{-2}$, while $\gamma\sim n$ for spin-1/2 wave. We thus have
$\tau_{\mbox{\tiny EM}}/\tau_{\mbox{\tiny S}} \sim n$,
which means that, in the $|n| \gg 1$ regime, the $Q$ value of the
electromagnetic cavity is also higher than that of the spin-1/2 Dirac cavity.

\begin{figure}
\centering
\includegraphics[width=\linewidth]{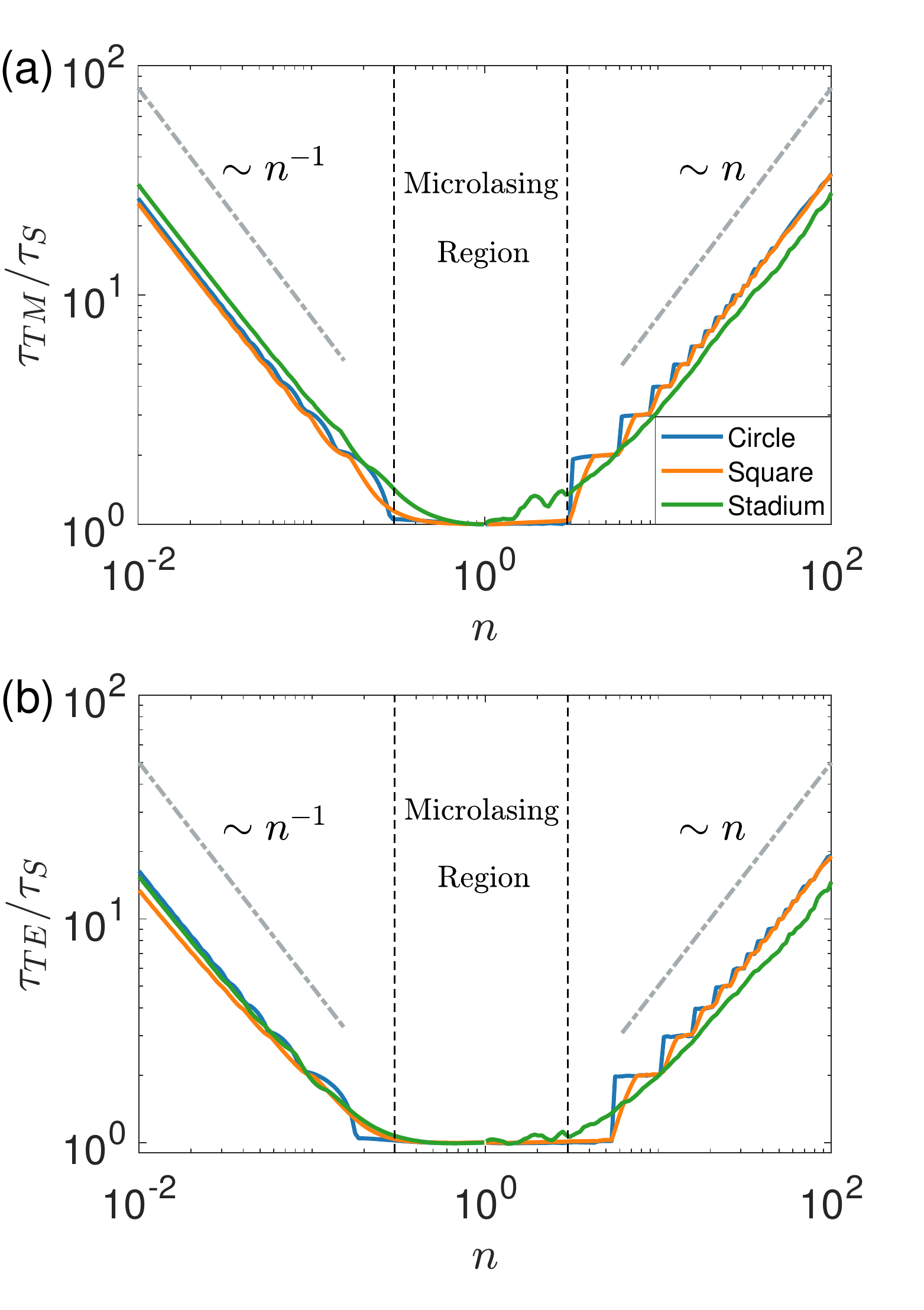}
\caption{ \textbf{Scaling with the relative refractive index of the ratio
between the mean escape time for an electromagnetic wave and that of a
spin-1/2 wave}: (a) the ratio $\tau_{\mbox{\tiny TM}}/\tau_{\mbox{\tiny S}}$
and (b) the ratio $\tau_{\mbox{\tiny TE}}/\tau_{\mbox{\tiny S}}$,
for three different types of cavity shapes (two with integrable and one with
chaotic ray dynamics). For $n \ll 1$, the ratio is proportional to
$n^{-1}$ while it is proportional to $n$ for $n \gg 1$, regardless of
whether the electromagnetic wave in comparison is TM or TE. The scaling laws
hold {\em regardless} of whether the classical ray dynamics are integrable
or chaotic. The surprising result is the existence of an interval in $n$
in which the spin-1/2 wave can have a high $Q$ value and chaos enabled
directional emission as for the electromagnetic wave.}
\label{fig:ratio}
\end{figure}

The analytic results can be summarized as
\begin{equation} \label{eq:three_general}
\begin{split}
\tau_{\mbox{\tiny EM}}/\tau_{\mbox{\tiny S}} =\alpha_1 n^{-1} \ \text{for} \ n \ll 1, \\
\tau_{\mbox{\tiny EM}}/\tau_{\mbox{\tiny S}} =\alpha_2 n \ \text{for} \ n \gg 1,
\end{split}
\end{equation}
where $\alpha_1$ and $\alpha_2$ are constants that depend on the geometric
shape of the cavity. We see that the integrable and chaotic cavities share
the same scaling law of the lifetime ratios with the refractive index.
Figures~\ref{fig:ratio}(a) and \ref{fig:ratio}(b) show the numerically
obtained ratios $\tau_{\mbox{\tiny TM}}/\tau_{\mbox{\tiny S}}$ and
$\tau_{\mbox{\tiny TE}}/\tau_{\mbox{\tiny S}}$ versus $n$, respectively.
There is a good agreement between the numerical results and those in
Eq.~(\ref{eq:three_general}).
The remarkable result is that, while the decay of the spin-1/2 wave is
significantly faster than that of the electromagnetic wave in both the
$n \ll 1$ and $n \gg 1$ regimes, there exists a sizable interval about
$n = 1$ in which the decay rates of the two types of systems are comparable,
as shown in Figs.~\ref{fig:ratio}(a,b). In this interval, high $Q$ values
can be achieved for spin-1/2 particles. The striking phenomenon is that the
ratios in this interval can be maintained at values close to one,
{\em regardless of the nature of the classical ray dynamics}.
That is, high $Q$ values can be achieved for spin-1/2 particles in a chaotic 
cavity to the same extent as for electromagnetic waves so as to ensure 
nonisotropic emission. 

\section{Discussion} \label{sec:discussion}

To summarize, motivated by the question of whether high $Q$ and directional
coherent emission are achievable for spin-1/2 particles, we investigated the
cavity decay problem with classically distinct dynamics in the semiclassical
regime. Previous work on microlasing of electromagnetic
waves~\cite{NSC:1994,MNCSC:1995,NSCGC:1996,NS:1997,GCNNSFSC:1998}
established that deformed chaotic cavities can meet the two key requirements
for microlasing: high $Q$ value and effective directional emission.
For spin-1/2 particles, confinement can be realized through an external
electric field. Our analysis and numerical results indicate that, for
a spin-1/2 wave cavity (e.g., made of graphene), there exists an experimentally
reasonable range of the applied electric potential in which the two
requirements can be met. For example, a chaotic graphene cavity can
simultaneously have a high $Q$ value and good emission directionality.

More specifically, we have analyzed the survival probability in both integrable
and chaotic cavities. For the integrable cavity, the decay is exponential in
the $|n| < 1$ regime. Significantly better confinement in the sense of
algebraic decay of the survival probability can be achieved for the integrable
cavity in the $|n| > 1$ regime. For larger $n$ values (when the Fermi energy 
is close to the potential), confinement is more robust. This result 
\sout{is consistent with that obtained through wave scattering
analysis~\cite{HSB:2013} and also}
agrees with experimental measurement on
circular potential confinement~\cite{Zhaoetal:2015} where high quality
confinement is achieved for high angular momentum modes. For the chaotic
cavity, the survival probability decays exponentially with time for all
possible $n$ values. We note, however, that the quantum regime
in which the scattering theory is applicable is not the semiclassical regime
treated in our work. In fact, in previous work on confinement of spin-1/2
fermions in graphene~\cite{KNG:2006,BTB:2009,SB:2011,HSB:2013,SB:2014}, the
relevant wave regime is not close to being semiclassical. To search for
regimes where high $Q$ and nonisotropic coherent emission are
possible for spin-1/2 particles, we obtain analytic
formulas to compare the average lifetime with that of an electromagnetic wave
in the same cavity. A striking result is that the behavior of the ratio of the
average lifetimes of the two types of waves versus $|n|$ is independent of the
nature of the underlying classical ray dynamics. For both $|n| \ll 1$ and
$|n| \gg 1$ regimes, there are scaling laws governing the ratio, which
indicates that the average trapping time of the electromagnetic wave is
significantly longer than that of the spin-1/2 wave, in accordance with
intuition. However, counter intuitively, there exists a regime of $|n|$ values
centered about one in which the average lifetimes for the two types of waves
are approximately the same, which is valid for both integrable and chaotic
cavities, generating remarkable decaying behavior of a spin-1/2 wave in this
regime.

We provide a brief discussion about the issue of directional emission. 
In optical microcavities, directional emission is typically shape dependent.
For example, in Sec.~VII of Ref.~[\onlinecite{CW:2015}], a number of 
high-Q cavities were described, which are able to emit light in certain
directions. For some specific cavity shape, it is possible to determine the 
probability of directional emission through ray tracing. For example, in 
Ref.~[\onlinecite{WH:2008}], a heart shaped cavity was studied, 
where the emission direction depends on some long lasting orbits with initial 
incident angle $|\sin \theta|<1/n$. As ray trajectories associated with
these orbits escape, radiation is 
generated but is concentrated in some special direction. In general, ray 
tracing is insufficient for determining if the underlying cavity can have 
directional emission. Instead, a wave approach based on scattering and 
solutions of the Dirac equation is necessary. Since a spin-1/2 system exhibits 
similar characteristics in the dependence of the transmission on the angle
to those of light (e.g., the transmission reaches a maximum for $\theta=0$ 
and a minimum for $\theta=\theta_c$), it is possible for high-Q operation 
in spin-1/2 systems to possess directional emission.

We would also like to explain the difference between the results from a 
relevant recent work~\cite{XWHL:2018} and those in the current work. 
Specifically, Ref.~[\onlinecite{XWHL:2018}] treated a scattering problem in 
graphene systems, where the spin degeneracy of the electrons is lifted through 
an exchange field from induced ferromagnetism. The scattering region has a 
non-concentric type of ring geometry, where a different gate potential is 
applied to the inner circle and to the region outside the inner circle but 
within the outer circle, respectively. As a result, the scattering dynamics 
for spin-up and spin-down electrons are characteristically different, both 
classically and quantum mechanically. For example, for proper values of the
gate potentials and eccentricity, the classical dynamics of spin-down 
electrons are completely integrable, while spin-up electrons exhibit fully 
developed chaotic scattering. Not only are the classical dynamics distinct, 
the corresponding quantum scattering also exhibits drastically different 
characteristics in terms of experimentally accessible quantities such as 
the cross sections, resonances, and the Wigner-Smith delay time. In 
Ref.~[\onlinecite{XWHL:2018}], the simultaneous coexistence of two different 
types of scattering behaviors is coined by the term ``relativistic quantum 
chimera.''

The focus of the current work is on optical like decay behaviors of Dirac
fermions in the semiclassical regime in the absence of any induced
ferromagnetism. There is then no splitting of the Dirac cone structure,
i.e., the energy bands of spin-up and spin-down electrons are completely
degenerate, ruling out the possibility of any relativistic quantum chimera
state. The results reported here on the decay of semiclassical massless
Dirac fermions from integrable or chaotic cavities thus do not depend on
the electron spin. Also note that, in Ref.~[\onlinecite{XWHL:2018}], while 
the classical dynamics were obtained using the same approach of Dirac 
electron optics as in the current work, the quantum scattering dynamics 
were calculated and analyzed based on solutions of the Dirac equation for 
two-component spinor waves.

\section*{Acknowledgement}

We would like to acknowledge support from the Vannevar Bush Faculty Fellowship
program sponsored by the Basic Research Office of the Assistant Secretary of
Defense for Research and Engineering and funded by the Office of Naval
Research through Grant No.~N00014-16-1-2828. DH is supported by the DoD
LUCI (Lab-University Collaborative Initiative) Program.

\appendix

\section{Survival probability of electromagnetic waves}
\label{Appendix_A}

For TM electromagnetic waves, the reflection coefficient is given
by~\cite{Jackson:book}
\begin{equation} \label{eq:B_R_TM}
R_{\mbox{\tiny TM}}(\theta) = \left( \frac{n\cos{\theta} - \cos{\theta_t}}
{n\cos{\theta}+\cos{\theta_t}} \right)^2,
\end{equation}
where $\theta$ and $\theta_t$ are the incident and refractive angles,
respectively, and $n$ is the relative refractive index. The formula for
TE waves is
\begin{equation} \label{eq:B_R_TE}
R_{\mbox{\tiny TE}}(\theta)= \left( \frac{\cos{\theta}-n\cos{\theta_t}}
{\cos{\theta} + n\cos{\theta_t}} \right)^2.
\end{equation}
The law of refraction is $\theta_t=\sin^{-1}{(n\sin{\theta})}$. For conventional
dielectric materials, we have $n>0$.

\paragraph*{Integrable cavity with $n<1$.}
In contrast to spin-1/2 waves where whispering gallery modes survive in
the cavity for the longest time [Eq.~\eqref{eq:2_circle_1_1_calculation}],
for TM and TE electromagnetic waves, such modes are along the diameter with
$\theta=0$. Substituting $\theta=0$ into Eqs.~\eqref{eq:B_R_TM} and
\eqref{eq:B_R_TE}, from Eq.~\eqref{eq:2_circle_1_1_calculation}, we obtain
the orbit length as $2\cos{\theta}$ and the exponential decay exponent
$\gamma$ as
\begin{equation}
\gamma_{\mbox{\tiny TM}}=\gamma_{\mbox{\tiny TE}}=\ln\left( \frac{1+n}{1-n}\right).
\end{equation}

\paragraph*{Integrable cavity with $n>1$.}
We expand Eq.~\eqref{eq:2_circle_1_inf_expansion} near $\theta_c$ to obtain
\begin{equation}
G(\theta)\approx\alpha x^{1/2}+O(x).
\end{equation}
where
\begin{equation}
\alpha_{\mbox{\tiny TM}}=\frac{2n\sqrt{2\sqrt{n^2-1}}}{n^2-1}.
\end{equation}
The survival probability for a TM wave is
\begin{equation}
P(t)=\frac{2\sqrt{n^2-1}}{\alpha_{\mbox{\tiny TM}}^2}t^{-2}.
\end{equation}
Similarly, we have, for a TE wave,
\begin{equation}
\alpha_{\mbox{\tiny TE}}=n^2\alpha_{\mbox{\tiny TM}}.
\end{equation}

\paragraph*{Chaotic stadium cavity with $n<1$.}
Using $R(n=0)=1$ for all $\theta$, we expand $\ln{(1-T)}$ in the small $n$
regime to obtain
\begin{equation}
T=4n\cos\theta+O(n^2).
\end{equation}
The integral in Eq.~\eqref{eq:2_stadium_integration} can be evaluated as
\begin{equation}
-\int_0^{\pi/2}\ln{R\cos{\theta}}d\theta
=\int_0^{\pi/2} 4n\cos^2{\theta}d\theta=n\pi.
\end{equation}

For a TE wave, the integral is
\begin{equation}
-\int_0^{\pi/2}\ln{R\cos{\theta}}d\theta
=\int_0^{\pi/2} 4n d\theta=2n\pi.
\end{equation}

\paragraph*{Chaotic stadium cavity with $n>1$.}
The result can be found in Ref.~[\onlinecite{RLKP:2006}]. For a TM wave, the
integral in Eq.~\eqref{eq:2_stadium_integration} is
\begin{equation}
-\int_0^{\sin^{-1}{(1/n)}} \cos{\theta}\ln{(1-R)} \approx \pi n^{-2}.
\end{equation}
For a TE wave, the integral is
\begin{equation}
-\int_0^{\sin^{-1}{(1/n)}} \cos{\theta}\ln{(1-R)} \approx 2\pi n^{-2}.
\end{equation}
The exponential decay exponents $\gamma$ for the two cases are given by
\begin{eqnarray}
\gamma_{\mbox{\tiny TM}} & = & \frac{2\pi^2 r_0}{\langle d\rangle L n^2}, \ \mbox{and} \\
\gamma_{\mbox{\tiny TE}} & = & \frac{4 \pi^2 r_0}{\langle d\rangle L n^2}.
\end{eqnarray}

\vspace*{0.2in}

\section{Survival probability of spin 1/2 waves in chaotic stadium cavity}
\label{Appendix_B}

\paragraph*{The case of $|n|<1$.}
For $n=0$, the reflection coefficient $R$ is
\begin{equation}
R=\frac{1-\cos{\theta}}{1+\cos{\theta}}.
\end{equation}
We thus have
\begin{equation}
-\ln R = 2\left(\frac{\cos^2{\theta}}{2} +\frac{\cos^4{\theta}}{4}
+\cdots \right).
\end{equation}
The ray density is given approximately by $\cos{\theta}$. Using Wallis'
integral~\cite{sebah2002introduction}
\begin{equation}
W_n=\int_0^{\pi/2} \cos^n{x} dx = \frac{\Gamma\left(\frac{n+1}{2} \right)
\Gamma\left(\frac{1}{2} \right)}{2\Gamma\left(\frac{n}{2}+1 \right)}.
\end{equation}
we can evaluate the integral in Eq.~\eqref{eq:2_stadium_integration} as
\begin{equation}
-\int \ln(R) \cos{\theta} d\theta=\sum_{n=1}^\infty \frac{W_{2n+1}}{n},
\end{equation}
where $\lim_{n\rightarrow \infty}W_n=0$, so the series converges.
For small $n$ values, we have
\begin{equation}
R\approx\frac{1-\cos{\theta}-n\sin^2{\theta}}{1+\cos{\theta}-n\sin^2{\theta}}.
\end{equation}
Rewriting this as $R=R_0 \left[1+nf(\theta) \right]$, we have the reflection
coefficient for $n=0$ as
\begin{equation}
\ln{R} = \ln{R_0} + n f(\theta).
\end{equation}
where
\begin{equation}
f(\theta)= \frac{n\sin^2{\theta}}{1+\cos{\theta}}
-\frac{n\sin^2{\theta}}{1-\cos{\theta}}.
\end{equation}
The decay exponent $\gamma$ can be determined through
\begin{equation}
-\int \cos\theta \ln R = \sum_{i=1}^\infty \frac{W_{2i+1}}{i} + \frac{\pi}{2}n.
\end{equation}

\paragraph*{The case of $|n|>1$.}
For $|n|>1$, rays with incident angle near the critical value for total
internal reflection dominate the long time behavior of the survival
probability. Near the critical angle where $T$ is about zero, we can expand
Eq.~\eqref{eq:2_stadium_integration} as
\begin{equation}
-\int_0^{\pi/2} \cos{\theta}\ln (1-T) d\theta =
\int_0^{\sin^{-1}{(1/n)}} T\cos{\theta} d\theta.
\end{equation}
For $n\rightarrow\infty$,  we let $p=n\sin\theta$ and expand the integrand
in terms of $1/n$ to obtain
\begin{equation}
\frac{1}{n}\int_0^1 \frac{2\sqrt{1-\frac{p^2}{n^2}}\sqrt{1-p^2}}
{1+\sqrt{1-\frac{p^2}{n^2}}\sqrt{1-p^2}-\frac{p^2}{n}} dp \approx
\frac{4-\pi}{n}.
\end{equation}
The result at the $n\rightarrow -\infty$ limit is the same. We obtain
\begin{equation}
\gamma=\frac{(8-2\pi)\pi R}{\langle d\rangle |n| L}.
\end{equation}


%
\end{document}